# The opportunity of the 2016 transit of Mercury for measuring the solar diameter and recommendations for the observation

Costantino Sigismondi

*(ICRA/Sapienza Università di Roma and IIS F. Caffè)*

**Abstract:** The transit of Mercury occurred two times in this century: 2003, May 7 and 2006, November 8. In 2016 there is another opportunity to observe this phenomenon and measure the solar diameter with the method of comparing the ephemerides with the observations. This method has been presented by I. I. Shapiro in 1980, the data of the observed transits (since 1631) have been re-analyzed by Sveshnikov (2002) and an improvement on the observed data, to avoid the confusion given by the black-drop effect, proposed by C. Sigismondi and collaborators since 2005 exploits the measurement of the chord drawn by the solar limb with the disk of the transiting planet and its extrapolation to zero through the analytic chord fitting the observations before the black drop, in the ingress/egress phases. A network of European observers (IOTA/ES) and observatories (coronograph of Bialkow, PL 56 cm; IRSOL, Locarno CH - 45 cm Gregorian telescope) is active for the 2016 transit. Recommendations to improve the observation of the ingress/egress phases and examples of the data analysis on the 1999 Mercury transit observed by TRACE are also done: the planet should be centered in the field of view of the telescope to avoid optical distortions, and the observations should be prolonged well after the black drop phase, estimated in about 40 seconds and more after the internal contact t2 in the ingress and before the internal contact t3 in egress.

## Introduction

On May 9, 2016 there is the fifth planetary transit of this century: 3 of Mercury (2003, 2006 [8] and 2016) and 2 of Venus (June 8, 2004 and June 6, 2012). For measuring the solar diameter Venus offered a unique opportunity with long lasting ingress and egress phases of almost 20 minutes, while for Mercury such phases last about 4 minutes; the grazing transit of 1999 was longer and gave the opportunity to study the black drop, similar to the shadow's attractions [4,10].

The comparison between the ephemerides and the observed timings of the ingress/egress allows us to compute the diameter of the Sun, in the hypothesis of spherical Sun [1,2,5].
This hypothesis is well confirmed within the errorbars of our measurements (RHESSI data, 2008 and SDS data show the Sun circular with an oblateness of 8 parts in $10^6$ [6]).
Dicke (1960s) searched for a solar oblateness to find room for his scalar-tensorial theory of gravitation alternative to Einstein's one. Now the interest in the solar figure is mainly for its subsurface activity and for the variations of the diameter with eventually consequences on Earth's climate [7].

**Method and data useful for measuring the solar diameter**
It is important to stress that the absolute timing of the four contacts is important, but it is more important to have several images (>50) with their absolute timings taken during the ingress (this is visibile in all Europe). From the images, as in the case of planetary images, we can select them with good local seeing, or temporary calm (being the seeing a statistical measurement of the optical turbulence of the atmosphere).
50 of such images sampling the 191 s of the ingress will allow to fit the chord planet-solar limb and to find the times when it is zero (t1 and t2 of the ingress). These two (absolute) times t1 external contact, t2 internal contact, will be compared with the ephemerides. The location (adress and city) of the observer has to be known to compute precise ephemerides.
**Summary: 1.** A sequence of UTC timed photos (or frames from a video) is recorded during the ingress phase (egress when available); **2.** the best photos of each stage are selected for avoiding seeing confusion; **3.** the analytic function of the chord c(t) [3] is fitted to the data to find c(t1, t2)=0. **4.** the comparison with the ephemerides of such t1, t2 allows us to calculate the variations of the solar radius with respect to the standard one of 959.63″ in visible light. Note that in case of Hα filters the value of the solar radius is different, so a note on the filters used during the transit for recording the photo is important.

As "best choice" method we select about 50 images like in fig. 1 during the ingress from 13:12:10 to 13:15:21 TMEC.

During the preparation to this event is crucial to verify how to focus perfectly the solar spots visible, and the maximum resolution obtainable with the "best choice" method.

**5.** In alternative to **3.**: the distance d(t) of the entering Mercury limb to the solar limb is a linear function of time, when it exceeds the angular diameter ±ΔD of Mercury, t2±Δt2 is found. F. Consalvi and L. Brescia (IIS Caffè, 1BT) made a graphical fit to the distances d(t) and Mercury diameter $D_M$ measured from Fig. 3, finding a Δt2=±57s, 1/31 of Mercury diameter, i.e. 0.4″.

**Recommendations for the observation**

The center of the field of view should be in the position where Mercury enters in the solar disk at the Position Angle 83°.2,[1] in the East side of the Sun. To find it leave the telescope without tracking, or hourly motion, and the direction where the image moves is West, 270°, the opposite East 90°. The solar spot which is currently (May 6[th]) on the solar disk (the number 100 in the Locarno Specola drawing)[2] is in the Northern side near the position angle of Mercury's ingress, to understand where the planet enters (above the line E-W parallel to the celestial equator). Because of the b0 angle[3] the point at PA=83° is below the solar equator, which is not parallel, for 9 may 2016, to a celestial parallel.

---

[1] http://eclipsewise.com/oh/tm2016.html
[2] http://www.specola.ch/drawings/2016/loc-d20160506.JPG
[3] http://www.petermeadows.com/html/location.html use Helioviewer soft.

Mercury should be in the center of the field of view because of the deformation of the field induced by the telescope.

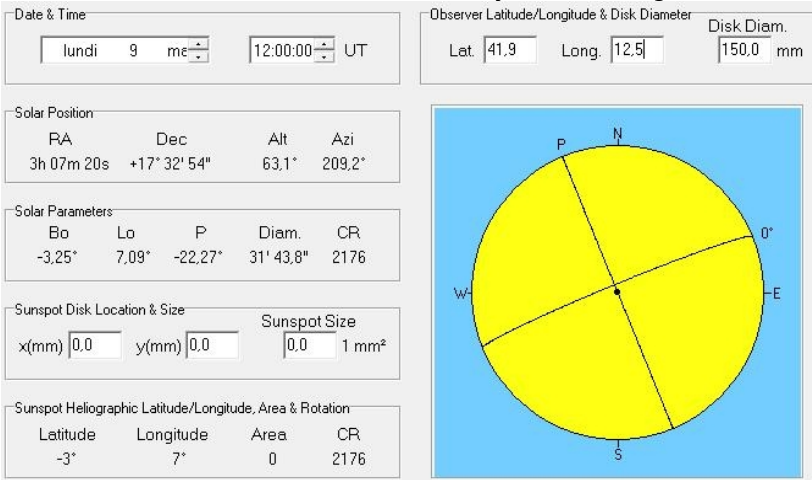

Fig. 1 Solar equator and celestial North (Helioviewer v. 3.2).

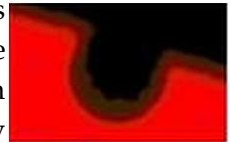

Fig. 2 The Hα image of the transit of Venus observed by Antony Aymamitis[4] on 8 June 2004 h 7:34 in Athens show this deformation problem: the Sun centered in the field of view and Venus at 15 arcminutes; the deformation of Venus radial/tangential is evident [11].

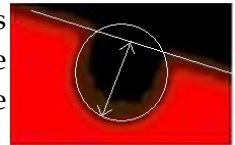

The alternative method **5.** of the previous point exploits the measurements from the image of the diameter of the planet and the distance limb-limb opposite, arrow in Fig. 3.

For this method a round planet's image is suitable, while for the chord method (perpendicular to it) direct data are used, without extrapolations of the solar limb position.

**The black drop duration**

An estimate for the black drop duration by simple proportion

---

4   C. Sigismondi, A. Ayomamitis, et al. ArXiv:1507.03622 (2015).

is made on the images of TRACE for the 1999 grazing transit.

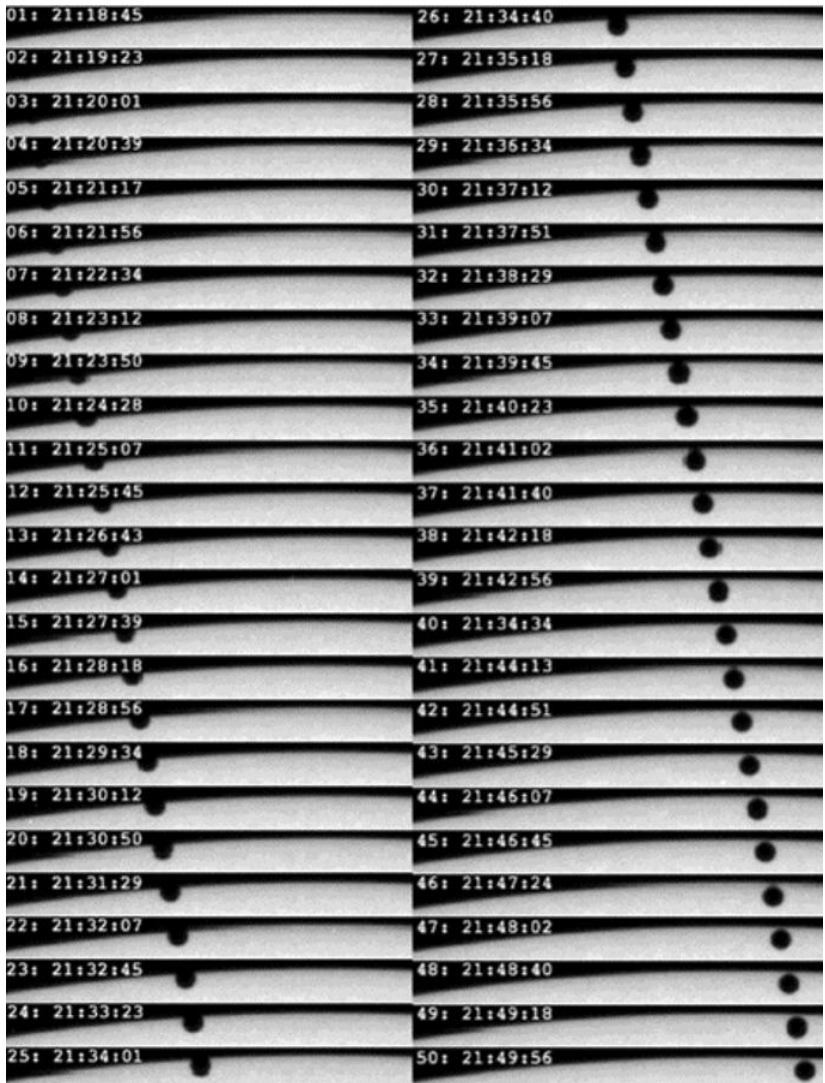

Fig. 4 Ingress of 1999 Mercury transit observed by TRACE.
TRACE, Transition Region and Coronal Explorer, satellite of NASA, observed the transit of Mercury of 15 November 1999

[4,9]: these images helped to explain the optical origin of the black drop. The phase of black drop appears in the last 10 images, lasting 1/5 of the ingress phase. In 2016 case the duration of the ingress is 191s, and 1/5 of it is 40s, as the duration of the black drop phase.

The telescope of TRACE has a diameter of 30 cm and its PSF is 1"; [5] a smaller telescope and subjected to the atmospheric blurring (not in the space) may experience longer black drop phase, as Wittmann [12] shows in its numerical simulation of the Mercury transit black drop phenomenon, and Guido Horn d'Arturo explains in 1922 including eye's astigmatism [13].

---

5   http://trace.lmsal.com/Project/Instrument/insrument.htm